\documentclass[reprint,prb,superscriptaddress,amsmath,amssymb,aps,floatfix
]{revtex4-2}

\usepackage{bm}%
\usepackage[colorlinks=true,linkcolor=blue]{hyperref}%
\expandafter\ifx\csname package@font\endcsname\relax\else
 \expandafter\expandafter
 \expandafter\usepackage
 \expandafter\expandafter
 \expandafter{\csname package@font\endcsname}%
\fi
\usepackage{graphicx}
\usepackage{dcolumn}
\usepackage{tabularx}
\makeatletter
\let\revtex@makecaption\@makecaption
\long\def\@makecaption#1#2{%
  \def\revtex@captype@figure{figure}%
  \ifx\@captype\revtex@captype@figure
    \par
    \vskip\abovecaptionskip
    \begingroup
      \small\rmfamily
      \samepage
      \flushing
      \let\footnote\@footnotemark@gobble
      \@make@capt@title{#1}{#2}\par
    \endgroup
    \vskip\belowcaptionskip
  \else
    \revtex@makecaption{#1}{#2}%
  \fi
}
\makeatother

\newcommand{\wmk}{W\,m$^{-1}$\,K$^{-1}$}
\newcommand{\cmm} {cm$^{-1}$}

\DeclareMathAlphabet{\mathsfit}{\encodingdefault}{\sfdefault}{m}{sl}
\SetMathAlphabet{\mathsfit}{bold}{\encodingdefault}{\sfdefault}{bx}{sl}
\newcommand{\tens}[1]{\bm{\mathsfit{#1}}}
\newcommand{\tenscomp}[1]{\mathsfit{#1}}

\begin{document}

\title{Anharmonic enhancement of thermal conductivity in amorphous Ge\texorpdfstring{$_2$Sb$_2$Te$_5$}{2Sb2Te5}: A consistent picture from molecular dynamics and Wigner transport}

\author{Luiz Felipe C. Pereira}
%\email{luiz.cpereira@ufpe.br}
\affiliation{Departamento de F\'{\i}sica, Centro de Ciências Exatas e da Natureza, Universidade Federal de Pernambuco, Recife 50670-901, Brazil}
\affiliation{Dipartimento di Fisica, Sapienza Università di Roma, Roma 00185, Italy}

\author{Kamil Iwanowski}
\affiliation{Department of Applied Physics and Applied Mathematics, Columbia University, New York, USA}

\author{Michele Simoncelli}
\affiliation{Department of Applied Physics and Applied Mathematics, Columbia University, New York, USA}

\author{Omar Abou El Kheir}
\affiliation{Department of Materials Science, University of Milano-Bicocca, Milano 20125, Italy}

\author{Marco Bernasconi}
\affiliation{Department of Materials Science, University of Milano-Bicocca, Milano 20125, Italy}

\author{Riccardo Mazzarello}
\affiliation{Dipartimento di Fisica, Sapienza Università di Roma, Roma 00185, Italy}

\author{Francesco Mauri}
\affiliation{Dipartimento di Fisica, Sapienza Università di Roma, Roma 00185, Italy}
 
\begin{abstract}
Anharmonicity limits heat conduction in crystals, but its role in amorphous solids, where heat is carried mainly by tunneling between quasi-degenerate vibrations, remains unsettled. Here we disentangle the contributions of structural disorder, quantum statistics, and anharmonicity to the thermal conductivity of amorphous Ge$_2$Sb$_2$Te$_5$ (aGST), the prototypical phase-change memory material. Using a single machine-learned potential trained on first-principles data, we compare classical equilibrium and nonequilibrium molecular dynamics (MD), which include anharmonicity to all orders, with harmonic Allen--Feldman theory and the anharmonic Wigner transport equation, both of which include quantum statistics. Four MD approaches agree on a conductivity of 0.171--0.181~\wmk{} at room temperature, about 16\% above the harmonic Allen--Feldman value for the same model, and the Wigner formulation bridges the two: with classical statistics, it agrees with MD to within 5\%. We trace this anharmonic enhancement to vibrational couplings that grow with the frequency difference between modes, so that anharmonic broadening opens additional channels for heat conduction. A comparison with silica, alumina, and disordered carbons reveals three regimes, in which anharmonicity reduces, barely affects, or enhances the conductivity. The predicted glass-like increase of the conductivity with temperature is consistent with thermoreflectance measurements. These results show that anharmonicity has no universal effect in disordered solids, and can promote rather than limit heat transport in strongly disordered materials.
\end{abstract}

\maketitle

\section{Introduction}

The thermal conductivity of a material is a key property that provides fundamental insights into its structure and scattering mechanisms, and determines its performance in thermal management and thermoelectric applications. 
Heat transport in solids can generally be described by the Fourier law, which relates a heat current to a temperature difference through a proportionality coefficient $\kappa$ called thermal conductivity,
\begin{equation}
{\bf j} = - \kappa \nabla T.
\label{eq01}
\end{equation}
In general, the thermal conductivity is a rank-2 tensor which reflects the symmetries (or lack thereof) of the material's atomic structure.

In dielectric crystals heat conduction is mediated by quantized lattice vibrations, i.e. phonons. The established phonon Boltzmann transport equation (BTE) describes how they carry heat as particle-like excitations, i.e., with velocities determined by their dispersion and over finite mean free paths determined by anharmonic interactions \cite{Ziman1960,broidoIntrinsicLatticeThermal2007}.
In contrast, in strongly disordered insulators the lack of periodicity blurs the dispersion relation, implying that heat transport at noncryogenic temperatures is mainly mediated by wave-like tunneling mechanisms between quasi-degenerate vibrational eigenstates first discussed by the harmonic theory developed by Allen and Feldman (AF) \cite{Allen1989,Allen1993}.
These mechanisms have been shown to emerge as special limits from a unified Wigner transport equation (WTE), which encompasses both propagation and tunneling mechanisms, also covering the intermediate regimes where both anharmonicity and disorder are relevant \cite{Simoncelli2019}. An equivalent method has been derived relying on the many-body Green's function formalism \cite{Isaeva2019,caldarelliManybodyGreensFunction2022}.

Several computational methods are currently available to predict the lattice thermal conductivity (LTC) of materials. They can be classified into two broad classes: those based on lattice dynamics (LD) calculations and those based on molecular dynamics (MD) simulations.

The first class comprises the BTE, AF, and WTE approaches, and naturally resolves the macroscopic thermal conductivity in terms of microscopic heat-transport mechanisms.
The particle-like heat transport mechanisms relevant to crystals can be successfully described by the BTE, which requires the knowledge of the phonon dispersion and anharmonic scattering operator determined by third- or higher-order derivatives of the Born--Oppenheimer potential energy surface \cite{paulattoAnharmonicPropertiesGeneralized2013,fengQuantumMechanicalPrediction2016,xiaFirstprinciplesTheoryFive2025}. It can be solved within the relaxation time approximation (RTA) or exactly through iterative \cite{omini1997heat}, variational \cite{Fugallo2013}, or exact-diagonalization \cite{chaputDirectSolutionLinearized2013,cepellottiThermalTransportCrystals2016} methods.
For amorphous solids, the Allen--Feldman theory provides an approach to determine the LTC in the harmonic approximation from the knowledge of the vibrational frequencies and related eigenvectors of the dynamical matrix \cite{Allen1989, Allen1993, Feldman1993}. Within this approach, the LTC is limited only by the structural disorder of the material and not by intrinsic anharmonicities.
The WTE applies to systems ranging from crystals to anharmonic amorphous solids, and its evaluation requires all the ingredients needed to evaluate the BTE and the AF model \cite{Simoncelli2019, Simoncelli2022, Simoncelli2023}.
In the case of BTE and WTE, anharmonic effects are computed perturbatively by expanding the Hamiltonian beyond the second (harmonic) order to account for phonon-phonon scattering. 

In the second class of methods based on MD, the atomic motion is simulated over a certain time interval, and the LTC is calculated from the resulting trajectories. MD methods can be further classified into equilibrium and nonequilibrium approaches. The former are based on linear response theory and do not require a finite temperature gradient, whereas the latter rely on the presence of a temperature gradient or a finite heat flux.
Concerning anharmonic effects, MD-based methods inherently account for all possible phonon-phonon scatterings relevant at the simulation temperature. However, in general, it is not possible to resolve these processes in terms of the number of phonons involved.

The description of atomic interactions lies at the heart of LD- and MD-based methods.
Until recently, LD-based approaches relied almost exclusively on ab initio forces to achieve high accuracy and were thus restricted to systems with small unit cells.
Conversely, MD-based methods require large supercells for statistical accuracy and were mostly limited to empirical potentials for computational efficiency.
This scenario changed dramatically with the advent of machine-learned interatomic potentials (MLIPs)\cite{Blank1995, Behler2007,batznerE3equivariantGraphNeural2022}, which enable simulations of thermal transport with the accuracy of ab initio methods and a computational efficiency close to that of empirical potentials\cite{sossoThermalTransportPhasechange2012,Lee2024,togoOntheflyTrainingPolynomial2024,potaThermalConductivityPredictions2025}.
With the use of MLIPs it is now possible to compute vibrational properties of systems with thousands of atoms, as well as simulate systems with millions of atoms with ab initio accuracy \cite{Zhou2025,omar2025}, all within reasonable computational resources.

In this work we investigate the LTC of amorphous Ge$_2$Sb$_2$Te$_5$ (aGST), a prominent phase-change material (PCM) used in optical storage devices and nonvolatile memories 
\cite{WUTTIG2007,zhang2019designing,noe2017phase,BoniardiRedaelli}.
These technologies exploit the pronounced contrast in optical reflectivity and electrical resistivity between the crystalline and amorphous states of the material, as well as the rapid and reversible transformation between the two states upon heating. 
More specifically, in memory devices, phase transitions are induced by applying suitable electrical pulses, which raise the temperature within PCM cells via the Joule effect. 
Therefore, thermal conductivity is a key property for device operation, as these processes strongly depend on heat dissipation and transport.
Previous works have investigated the thermal conductivity of aGST using Density Functional Theory (DFT) \cite{Duong2021,massobrio2,Yang2021} and \textit{ab initio} approach-to-equilibrium MD \cite{Mocanu2020,Lampin2013}. However, noticeable discrepancies still persist among the reported literature values. In this work, we take a step further by systematically comparing results from LD and MD methods to assess their consistency and identify the contribution of anharmonicity to the LTC of aGST.

We employ equilibrium and nonequilibrium MD simulations, the Allen--Feldman model, and the WTE, with interatomic forces modeled by the neuroevolution potential (NEP) \cite{Fan2021}.
This MLIP is trained on the DFT database of Ref.~\cite{ElKheir2024}, computed with the Perdew--Burke--Ernzerhof \cite{Perdew1996} exchange and correlation functional.
The LTC of aGST is obtained from equilibrium MD using both the Green--Kubo and Helfand--Einstein formalisms.
Additionally, the reverse nonequilibrium MD (RNEMD) and homogeneous nonequilibrium MD (HNEMD) approaches yield results consistent with the equilibrium simulations.
The LTC of aGST is further evaluated within the fully harmonic approximation using the Allen--Feldman model, and up to the third-order anharmonic case using the WTE. Higher-order anharmonic effects are assessed via the MD simulations.
We find excellent agreement between MD- and WTE-based results, confirming the validity of all approaches within their respective domains of validity.
Remarkably, our results show that anharmonicity enhances the thermal conductivity of aGST, in contrast to other amorphous solids such as silica and alumina \cite{Simoncelli2023, Harper2024}.

\section {The neuroevolution potential}

NEP is a machine-learned interatomic potential implemented in the open-source package GPUMD \cite{Fan2017, Fan2021}.
It uses a feedforward neural network with a single hidden layer to represent the energy of an atom as a function of a descriptor vector.
The local atomic environment is described using a set of radial and angular basis functions constructed analytically from Chebyshev and Legendre polynomials.
It is named after the separable natural evolution strategy, an efficient and robust optimization method used to train the neural network parameters.
A principal advantage of NEP is its computational efficiency, achieved through a native implementation on graphics processing units (GPUs).

\begin{figure}
\includegraphics[width=\linewidth]{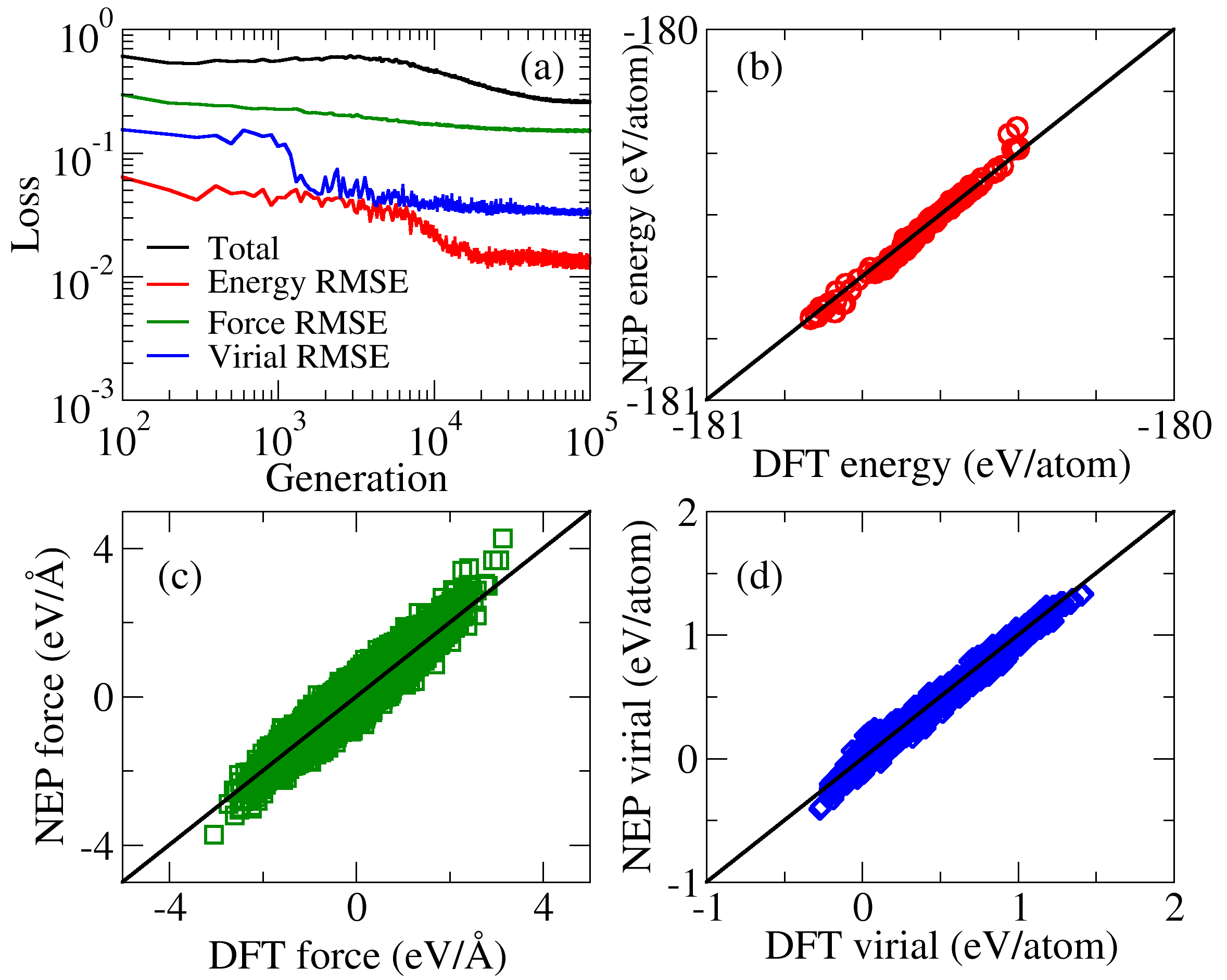}
\caption{(a) Total loss function, energy root-mean-square error (RMSE), force RMSE, and virial RMSE during training of NEP. (b)–(d) Comparison
between NEP predictions and DFT reference data for energy,
force, and virial components. Straight 
lines in panels (b)–(d) indicate a perfect agreement.}
\label{fig:train-nep}
\end{figure}

The training set is the full DFT database of Ref.~\cite{ElKheir2024}, which comprises liquid, supercooled-liquid, amorphous, cubic, and hexagonal configurations of GST. The reference energies, forces, and virials were computed with CP2K using the PBE functional without dispersion corrections, Goedecker--Teter--Hutter norm-conserving pseudopotentials, a 400~Ry density cutoff, and Brillouin-zone sampling on a $\bm{k}$-point mesh with Fermi--Dirac smearing at 300~K \cite{ElKheir2024}.
We employ the NEP4 version of the potential \cite{Fan2022GPUMD,Xu2025GPUMD4}, trained with a radial cutoff of 7.0~\AA{} and an angular cutoff of 6.0~\AA, radial and angular descriptors with $n_{\rm max}=4$ and a basis size of 12, angular expansions up to $l_{\rm max}=4$ (three-body) and $l_{\rm max}=2$ (four-body), a hidden layer with 30 neurons, and $10^5$ generations of the separable natural evolution strategy.
All configurations were used for training. 
The quality of the trained potential is summarized in Fig.~\ref{fig:train-nep}, which compares NEP predictions with the DFT reference energies, forces, and virials used during the optimization.
The resulting root-mean-squared errors (RMSE) with respect to the DFT training data are 12.9~meV/atom for energies, 151~meV/\AA{} for forces, and 32.6~meV/atom for virials.

\section{Methods}

\subsection{Equilibrium molecular dynamics}

The fundamental quantity in the calculation of the thermal conductivity in MD-based methods is the heat flux.
Although its formulation is not unique and there is some freedom in its definition, any correct formulation should provide equivalent results \cite{Fan2015,Grasselli2021a}.
In the presence of many-body interactions the heat current can be conveniently written as
\begin{equation}
{\bf J}(t) \equiv \frac{d}{dt} \sum_i {\bf r}_i E_i = \sum_i {\bf v}_i E_i + \sum_i {\bf r}_i \frac{dE_i}{dt},
\end{equation}
where ${\bf r}_i$, ${\bf v}_i$ and $E_i$ are the position, velocity and total energy of atom $i$.
The actual heat flux is given by ${\bf j} = {\bf J}/V$.
The term involving the time derivative of the total atomic energy can be written in terms of the atomic virial, such that \cite{Fan2015} 
\begin{equation}
{\bf J}(t) = \sum_i {\bf v}_i E_i +  \sum_i \sum_{j \ne i} {\bf r}_{ij} \left( \frac{\partial U_j}{\partial {\bf r}_{ji}} \cdot {\bf v}_i  \right),
\end{equation}
where $U_j$ is the potential energy of atom $j$.
The first term on the right-hand side has a convective character; in solids with a stationary center of mass and negligible atomic diffusion, its contribution to the thermal conductivity is negligible.

The Green--Kubo formula relates transport coefficients to the integral of the autocorrelation function of the associated fluxes \cite{Kubo1957, Zwanzig1965}.
In the case of the lattice thermal conductivity, each component of the conductivity tensor can be written as
\begin{equation}
\kappa_{\alpha \beta} = \lim_{t,V \to \infty} \frac{1}{k_B V T^2} \int_0^t \langle J_{\alpha}(\tau) J_{\beta}(0) \rangle d\tau,
\end{equation}
where $k_B$ is Boltzmann's constant, $V$ is the system volume and $T$ its temperature.
The expression is strictly valid in the thermodynamic limit, i.e. for large enough systems, and for long enough integration times.

The general meaning of the Green--Kubo integral is the following: in equilibrium the average heat flux vanishes, and its correlation function decays to zero after an interval that is directly related to $\kappa$.
In crystalline systems the autocorrelation function decays almost monotonically, and the integral converges faster for materials with low conductivity and slower otherwise.
However, in amorphous systems the decay of the autocorrelation function is not monotonic and since they tend to have low thermal conductivities, it can be a challenge to assess convergence of the Green--Kubo integral.

An alternative way of computing the components of the conductivity tensor is the so-called Helfand--Einstein integral \cite{Helfand1960}, which relates $\kappa$ to the time derivative of the mean squared displacement of the integrated heat flux \cite{Allen1989a}.
This is analogous to computing the diffusion coefficient from the mean square displacement of the particles instead of their velocity autocorrelation function, as originally proposed by Einstein.
The method was adapted to heat transport by Helfand, and provides a better signal-to-noise ratio due to its mathematical construction \cite{Grasselli2021a}.
The Helfand--Einstein expression can be written as
\begin{equation}
\begin{split}
\kappa_{\alpha \beta} ={}& \lim_{t,V \to \infty} \frac{1}{V k_B T^2} \frac{1}{2} \\ 
&\times \frac{d}{dt} \left \langle \left[ \epsilon_\alpha(t) - \epsilon_\alpha(0) \right]\left[ \epsilon_\beta(t) - \epsilon_\beta(0) \right] \right \rangle,
\end{split}
\end{equation}
where $\epsilon_{\alpha}(t)=\int_0^t J_{\alpha}(t')dt'$, and the angle brackets indicate an average taken over independent origins, analogous to what is used in the representation of the autocorrelation function.

\subsection{Nonequilibrium molecular dynamics}

Nonequilibrium molecular dynamics methods can be employed to compute the thermal conductivity in the presence of a temperature gradient, such that the average heat flux does not vanish.
In the so-called direct method a temperature gradient is imposed in the system by connecting thermostats to spatially separated regions while the heat flux between such regions is measured.
Once the system is in a steady state, i.e. the temperature gradient and the heat flux reach stationary values, $\kappa_{\alpha \beta}$ is calculated directly from Fourier law:
\begin{equation}
\kappa_{\alpha \beta} = \lim_{t,V \to \infty} \frac{1}{V} \frac{\langle J_\alpha(t) \rangle_{\text{ne}}}{\nabla_\beta T}.
\end{equation}

It is also possible to invert the cause and effect relationship between temperature gradient and heat flux, as in the reverse NEMD (RNEMD) method \cite{Muller-Plathe1997}.
In this case, a heat flux is imposed in the system by swapping particles with small kinetic energy in one region with particles of high kinetic energy in another (spatially separated) region.
By measuring the temperature in chunks between the two regions it is possible to determine the temperature gradient, and thus, in the steady state regime, calculate $\kappa_{\alpha \beta}$ from Fourier law, as above.

A common aspect of both NEMD methods described above is that the final result is essentially independent of the initial conditions.
In other words, the variance among independent runs is negligible.
Thus one only needs to ensure that the system has reached the steady state where the average heat flux and average temperature gradient are stationary.
However, the transient time required to reach the steady state increases with the system size, such that larger systems require longer runs.

Another fundamental aspect of NEMD methods is the size-dependence, or rather, the dependence of the calculated conductivity on the length along the heat flux direction.
Due to the presence of a heat source and sink, there is an intrinsic limit to the maximum distance heat can propagate in the system, which is determined by the separation between heat reservoirs.
Therefore, heat carriers with a typical propagating length longer than the separation between the reservoirs will not propagate in the system, decreasing its thermal conductivity relative to the diffusive value \cite{Schelling2002, Felix2018, Pereira2021}.

\subsection{Homogeneous nonequilibrium molecular dynamics}

Originally proposed by Evans \cite{Evans1982} and later extended by Fan et al. \cite{Fan2019}, the HNEMD method presents a series of advantages over traditional equilibrium and nonequilibrium methods.
In this method the equations of motion are modified by the addition of a small, fictitious external force which drives the system out of equilibrium and generates a nonvanishing heat flux but no temperature gradient.
The LTC is then proportional to the average nonequilibrium heat flux and inversely proportional to the external force
\begin{equation}
\kappa_{\alpha \beta} = \lim_{t,V \to \infty} \frac{1}{V} \frac{\langle J_\alpha(t) \rangle_{\text{ne}}}{TF_{e,\beta}},
\end{equation}
where $F_{e,\beta}$ is the external force parameter which has units of inverse length, and should be just large enough to keep the noise-to-signal ratio in $\langle J_\alpha(t) \rangle_{\text{ne}}$ low.

\subsection{Anharmonic Wigner conductivity \& its special harmonic Allen--Feldman limit}

The Wigner transport equation (WTE) is a quantum kinetic equation for lattice vibrations that unifies and extends the Boltzmann and Allen--Feldman descriptions. It accounts, within the same framework, for quantum Bose--Einstein statistics, anharmonicity, and structural disorder, and therefore describes both particle-like propagation and wave-like tunneling of vibrational energy. In this formulation, transport follows from the evolution of the Wigner quasiprobability distribution in phase space, obtained from the Wigner--Weyl transform of the vibrational density matrix.
For finite-size atomistic models of amorphous solids, the Wigner formulation gives the following regularized WTE (`rWTE') conductivity\cite{Simoncelli2023}:
\begin{equation}
\begin{split}
\kappa{=}\frac{1}{{V}{N_{\rm c}} }{\sum_{\bm{q},s,s'}}&\!
\frac{\omega_{\bm{q}s}{+}\omega_{\bm{q}s'}}{4}\!\left(\frac{C_{\bm{q}s}}{\omega_{\bm{q}s}}{+}\frac{C_{\bm{q}s'}}{\omega_{\bm{q}s'}}\right)\!
\frac{|\!|\tens{v}(\bm{q})_{s,s'}|\!|^2}{3}\\
\times&\pi\mathcal{F}_{[\eta;\Gamma_{\bm{q}s}{+}\Gamma_{\bm{q}s'}]}(\omega_{\bm{q}s}-\omega_{\bm{q}s'})\;,
\label{eq:thermal_conductivity_combined}
\end{split}
\raisetag{5mm}
\end{equation}
Here, $\hbar\omega_{\bm{q}s}$ is the energy of vibration $\bm{q}s$, while $\hbar\Gamma_{\bm{q}s}$ is its linewidth, i.e., its energy uncertainty due to anharmonicity~\cite{paulattoAnharmonicPropertiesGeneralized2013,Fugallo2013,Li2014} and isotopic-mass disorder~\cite{tamuraIsotopeScatteringDispersive1983};
$|\!|\tens{v}(\bm{q})_{s,s'}|\!|^2{=}\sum_{\alpha=1}^3\tenscomp{v}^{\alpha}(\bm{q})_{s,s'}\tenscomp{v}^{\alpha}(\bm{q})_{s',s}$
is the squared modulus of the velocity operator \cite{Simoncelli2022} between eigenstates $s$ and $s'$ at fixed $\bm{q}$. The index $\alpha$ denotes a Cartesian direction; since amorphous solids are in general isotropic, the scalar conductivity in Eq.~(\ref{eq:thermal_conductivity_combined}) is obtained from the trace average of $\kappa_{\alpha\beta}$ \cite{Simoncelli2023}.
\begin{equation}
C_{\bm{q}s}{=}C[\omega_{\bm{q}s},T]{=}\frac{\hbar^2\omega^2_{\bm{q}s} }{k_{\rm B} T^2} {\tenscomp{N}}_{\bm{q}s}\big({\tenscomp{N}}_{\bm{q}s}{+}1\big)  
\label{eq:quantum_specific_heat_A}
\end{equation}
is the quantum specific heat of mode $\bm{q}s$.
Vibrational modes are labeled by both wavevector $\bm{q}$ and mode index $s$ because amorphous solids are represented by periodic cells containing a large but finite number of atoms $N_{\rm at}$.
The corresponding Brillouin zone (BZ) has a small but finite volume; the ideal-glass description is recovered in the limit of thermodynamically large simulation-cell volume ${V}$, which corresponds to a BZ reducing to $\bm{q}{=}\bm{0}$ only.
$N_{\rm c}$ and $\tenscomp{N}_{\bm{q}s}$ denote, respectively, the normalization factor that accounts for the number of $\bm{q}$ points used to sample the BZ, and the Bose--Einstein distribution.

When the structural-disorder length scale is shorter than the simulation cell, Fourier interpolation over this small BZ can improve the sampling of vibrational properties and enable an extrapolation of the bulk thermal conductivity \cite{Simoncelli2023}.
For a disordered model, this interpolation is equivalent to averaging over boundary conditions. A calculation at $\bm{q}{=}\bm{0}$ imposes periodic boundary conditions on the simulation box; a calculation at $\bm{q}{=}(1, 0, 0)\pi/L$ imposes antiperiodic boundary conditions along $x$ and periodic boundary conditions along the remaining directions (\citet{Feldman1993}).
Upon increasing the size of the simulation, the sensitivity to these boundary conditions must decrease. This convergence can be checked by comparing a $\bm{q}{=}\bm{0}$ calculation with one using Fourier interpolation; when the two agree, the system is large enough for a direct finite-cell estimate of the bulk limit.

The last factor in Eq.~(\ref{eq:thermal_conductivity_combined}), $\mathcal{F}_{[\eta;\Gamma_{\bm{q}s}{+}\Gamma_{\bm{q}s'}]}(\omega_{\bm{q}s}-\omega_{\bm{q}s'})$, is a Voigt distribution centered on the frequency mismatch between two vibrations. It reduces to a Lorentzian with full width at half maximum (FWHM) $\Gamma_{\bm{q}s}{+}\Gamma_{\bm{q}s'}$ when the anharmonic linewidth dominates the numerical broadening, $\Gamma_{\bm{q}s}{+}\Gamma_{\bm{q}s'}\gg \eta$,
\begin{equation}
\begin{split}
    &\mathcal{F}_{[\Gamma(\mathbf{q})_s{+}\Gamma(\mathbf{q})_{s'}]}\big(\omega(\mathbf{q})_s{-}\omega(\mathbf{q})_{s'}\big)\\
    &=
  \frac{1}{\pi}\frac{\frac{1}{2}\big(\Gamma(\mathbf{q})_s+\Gamma(\mathbf{q})_{s'}\big)}{\big(\omega(\mathbf{q})_s-\omega(\mathbf{q})_{s'}\big)^2+\frac{1}{4}\big(\Gamma(\mathbf{q})_s+\Gamma(\mathbf{q})_{s'}\big)^2};
\end{split}
  \label{eq:Lorentzian}
\end{equation}
whereas in the opposite limit, $\Gamma_{\bm{q}s}{+}\Gamma_{\bm{q}s'}\ll \eta$, it becomes a Gaussian representation of the Dirac delta with variance $\eta^2\pi/2$:
\begin{equation}
\lim_{\eta{\to}0}\!\!\left[\lim_{\mathcal{V}\to \infty}\mathcal{F}_{[2\eta]}(\omega(\mathbf{q})_s{-}\omega(\mathbf{q})_{s'})\right]
   {=}\delta\left({\omega(\mathbf{q})_s{-}\omega(\mathbf{q})_{s'} }\right).
  \label{eq:Gaussian}
\end{equation}

As discussed in Ref.~\cite{Simoncelli2023}, Eq.~(\ref{eq:thermal_conductivity_combined}) rationalizes heat transport in terms of couplings between pairs of vibrations whose energy difference is smaller than the width of the Voigt profile. In the low-temperature harmonic limit, the anharmonic linewidths vanish, $\Gamma_{\bm{q}s}{+}\Gamma_{\bm{q}s'}\to 0$, and the width is set by $\eta$. Choosing $\eta$ slightly larger than the average level spacing $\hbar\Delta\omega_{\rm avg}=\frac{\hbar\omega_{\rm max}}{3\cdot N_{\rm at}}$ ($\hbar\omega_{\rm max}$ is the maximum vibrational energy, while $3N_{\rm at}$ is the number of vibrational energy levels in the reference cell) preserves the wave-like tunneling between neighboring, quasi-degenerate vibrational eigenstates that survives when anharmonicity vanishes\cite{Harper2024}. In this limit, Eq.~(\ref{eq:thermal_conductivity_combined}) reduces to the harmonic Allen--Feldman (AF) thermal conductivity of glasses \cite{Allen1989}.
In contrast, when temperature causes the anharmonic linewidths to be much larger than the computational broadening $\eta$, the Voigt profile is controlled by the Lorentzian width $\Gamma_{\bm{q}s}{+}\Gamma_{\bm{q}s'}$, and Eq.~(\ref{eq:thermal_conductivity_combined}) reduces to the anharmonic Wigner conductivity expression~\cite{Simoncelli2022}.
Thus, for finite-size atomistic models of amorphous solids at temperatures for which $\Delta\omega_{\rm avg}> \Gamma_{\bm{q}s}$, the rWTE expression~(\ref{eq:thermal_conductivity_combined}) retains the correct low-temperature AF limit and introduces anharmonicity only when the linewidths are not dominated by finite-size level spacing \cite{Simoncelli2022}.

Brute-force convergence with system size is often prohibitively expensive. In the next section we show that, for finite-size models of strongly disordered aGST, the bulk conductivity can indeed be computed with rWTE, Eq.~(\ref{eq:thermal_conductivity_combined}). This strategy has been validated in Refs.~\cite{Simoncelli2023,Harper2024,iwanowski_bond-network_2025}: the method preserves tunneling between neighboring eigenstates, a property that otherwise emerges only in the thermodynamic limit, and improves vibrational sampling by averaging over boundary conditions.
We conclude by noting that the Wigner thermal conductivity expression~(\ref{eq:thermal_conductivity_combined}) can also be derived from a many-body Green--Kubo approach~\cite{Isaeva2019,caldarelliManybodyGreensFunction2022}.

\section{Results}

\subsection{Computational details}

In the present study, aGST supercells of cubic shape and varying size were prepared by quenching from the melt during multistep MD simulations.
The protocol consisted of a 10 ps run at 2000 K, a quench to 990 K in 50 ps followed by 10 ps of equilibration, and a final quench from 990 K to 300 K in 100 ps followed by 20 ps of equilibration at 300 K. 
The density was adjusted to the experimental room-temperature value of 5.85 g/cm$^3$. 
The amorphous models are characterized by their total number of atoms and nominal volume: 81 atoms (14 \AA)$^3$, 216 atoms (19 \AA)$^3$, 999 atoms (32 \AA)$^3$, and 1998 atoms (40 \AA)$^3$.
EMD and RNEMD simulations were performed with LAMMPS \cite{Plimpton1995} through the NEP\_CPU interface \cite{Xu2025GPUMD4}, while HNEMD simulations were performed with GPUMD \cite{Fan2013,Fan2017}.
No thermostat was employed during EMD simulations, while the Nosé--Hoover thermostat with chains \cite{Martyna1992} was applied throughout the RNEMD simulations, including the production stage, and the Bussi--Donadio--Parrinello stochastic thermostat \cite{Bussi2007} was employed during HNEMD.
An integration timestep of $1.0$ fs was used in all MD simulations.

Harmonic phonon calculations were performed with phonopy \cite{Togo2015}.
Atomic forces were minimized with a tolerance of $10^{-6}$ eV/\AA~ at fixed volume, with a combination of the conjugate gradient and the FIRE algorithms in LAMMPS.
Anharmonic third-order phonon calculations were performed with phono3py \cite{Togo2023}, with cutoff distances considering up to 7 nearest neighbors.
Second- and third-order force constants were computed with $0.01$ and $0.03$ \AA~ finite atomic displacements in LAMMPS, respectively.
Phonon lifetimes were computed with phono3py in the RTA, and the Brillouin zone integration was performed with the tetrahedron method for the 81-atom cell, while a Gaussian smearing function with $\sigma=0.06$ THz and a cutoff at 2.5 standard deviations was used for the 216-atom cell.

\subsection{Equilibrium molecular dynamics}

We begin by determining the LTC of aGST at $T=300$ K with the Green--Kubo integral and the equivalent Helfand--Einstein expression.
To achieve long-time convergence, we performed simulations spanning 90 ns, for a total of 90 $\times 10^6$ MD steps.
The heat flux was recorded every 10 fs for the calculation of the heat-flux autocorrelation function, or of the analogous mean squared displacement; both analyses used a maximum correlation time equal to 10\% of the total run time, i.e. 9.0 ns.

The running thermal conductivities are averaged over the three independent Cartesian coordinates as well as 4 independent runs, i.e. $\kappa(t)=\langle [\kappa_{xx}(t)+\kappa_{yy}(t)+\kappa_{zz}(t)]/3 \rangle$.
In spite of this averaging process, the data remains rather noisy, so we resort to a moving window average of the final $\kappa(t)$ data as proposed by Knoop, Scheffler, and Carbogno in Ref.~\cite{Knoop2023}.
We employ a window with a 2 ps width, equivalent to the inverse of the frequency corresponding to the first significant peak in the vibrational density of states (vDOS) of aGST.
Finally, we discard the first 1 ns from the running average due to excessive noise, such that the reported thermal conductivity with each method is the average $\kappa(t)$ in the final 8 ns, and the uncertainty is taken as the standard error of the mean in this interval.

\begin{figure*}
\includegraphics[width=0.9\linewidth]{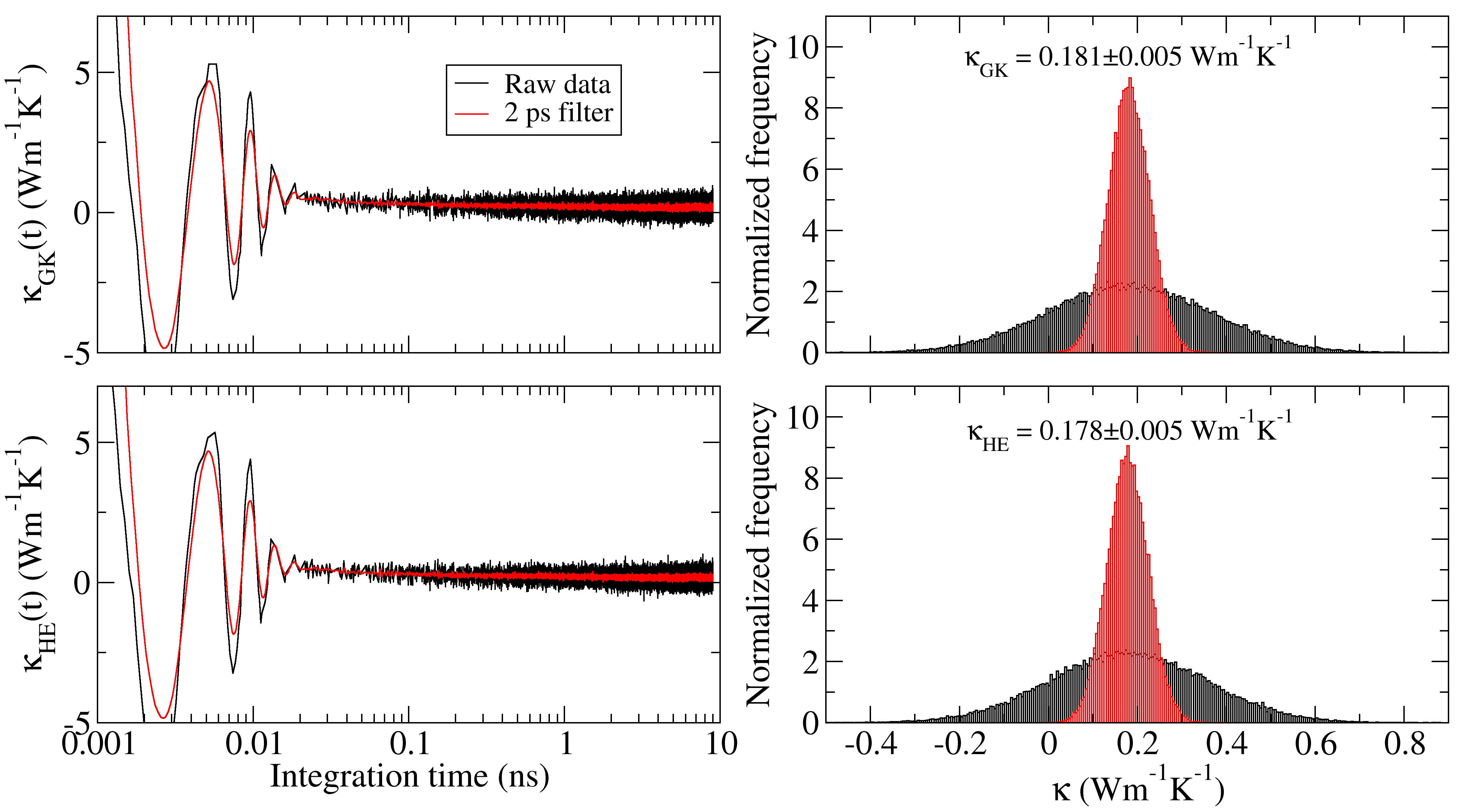}
\caption{Average running thermal conductivity computed with the Green--Kubo integral and the equivalent Helfand--Einstein expression for a 1998-atom cell at $T=300$ K. A moving average with a 2 ps window was applied to filter out excessive noise and decrease the variance; the resulting distributions are shown in the histograms.}
\label{fig:gk-he}
\end{figure*}

Figure \ref{fig:gk-he} presents the average running thermal conductivities computed with both methods.
In many cases, the Helfand--Einstein expression tends to result in a smaller variance when compared to the Green--Kubo method, but this is not the case in aGST.
The black and red curves represent the raw data and the filtered data, respectively, averaged over a 2 ps window, as mentioned above.
The noise reduction is apparent in the histograms shown in Fig.~\ref{fig:gk-he}, although the average does not change significantly.
For the 1998-atom system with nominal volume $V=(40$ \AA$)^3$, we obtain $\kappa_{\text{GK}}=0.181 \pm 0.005$ \wmk, and $\kappa_{\text{HE}}=0.178 \pm 0.005$ \wmk.

\subsection{Nonequilibrium molecular dynamics}

\begin{figure}
\includegraphics[width=\linewidth]{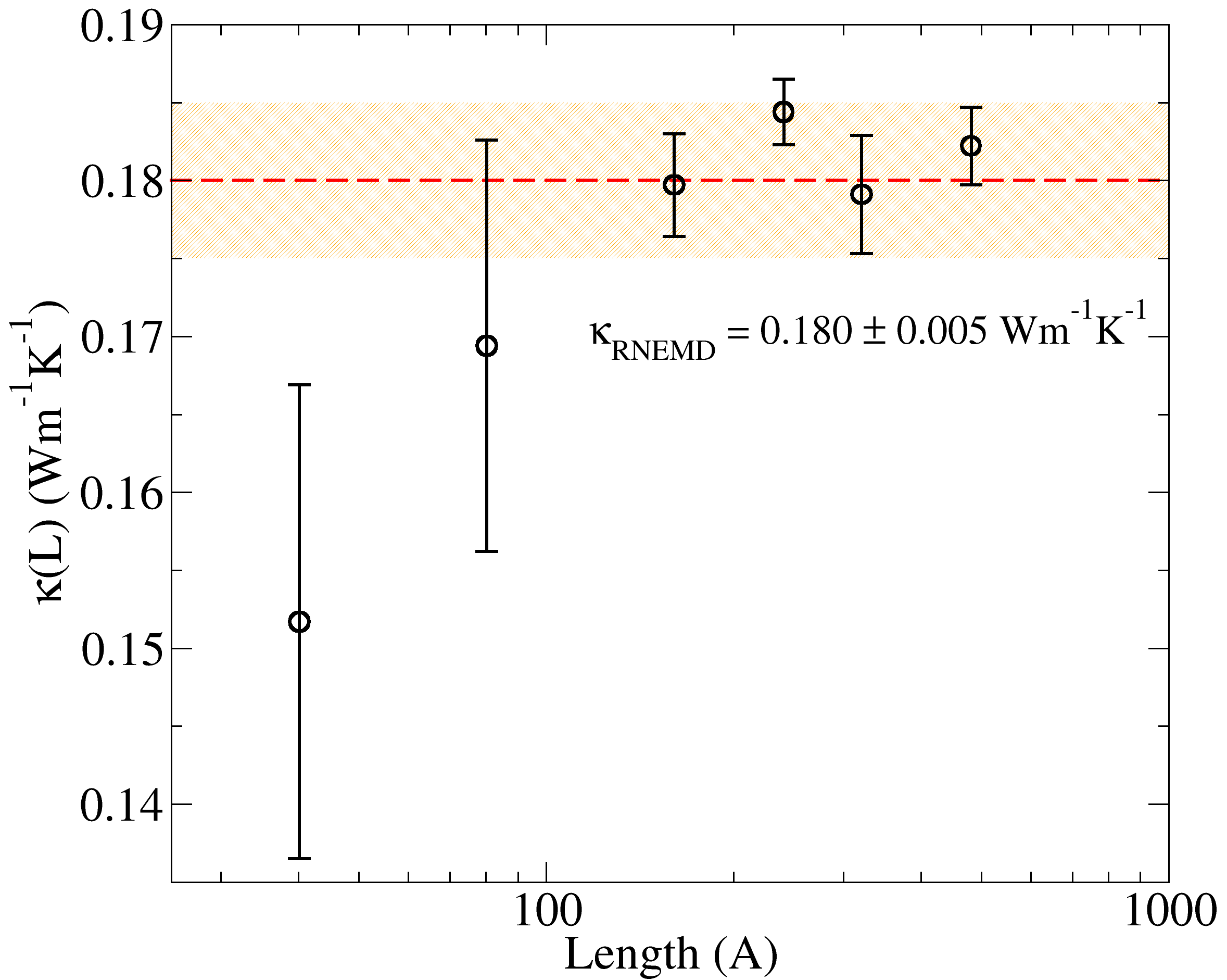}
\caption{Lattice thermal conductivity of aGST computed with the reverse nonequilibrium molecular dynamics method at $T=300$ K as a function of length. Each sample length was obtained by replicating the 1998-atom cell along the heat flux direction up to 12 times.}
\label{fig:rnemd}
\end{figure}

To determine convergence with length along the heat-transport direction, we calculate the LTC of simulation cells with increasing size along the heat-flux direction.
Those cells are generated by replicating the original supercell an integer number of times along one of the Cartesian directions, and their length $L$ refers to the full length of the simulation cell along the heat-flux direction.
Figure \ref{fig:rnemd} shows the results obtained replicating the 1998-atom cell up to 12 times, for a maximum of 23,976 atoms, and performing the kinetic energy swaps every 500 MD steps.
The error bars are calculated by propagating the statistical uncertainties in the average heat flux and the average temperature gradient, both in the stationary regime.
Within RNEMD, the time required to reach the steady state tends to increase with the system length.
In this case, we performed 4 ns runs for the two smallest cells, 8 ns for the following two, and 12 ns for the two largest cells.

The data in Fig.~\ref{fig:rnemd} indicates diffusive heat transport for systems longer than $\sim 100$ \AA, so there is no need to perform any heuristic fit to extrapolate from the ballistic to the diffusive regime. 
In this case, we obtain $\kappa_{\text{RNEMD}}=0.180 \pm 0.005$ \wmk, in agreement with EMD results but smaller than previous ab initio studies which obtained  $\kappa=0.25 \pm 0.01$ \wmk \cite{Duong2021}.
Since the calculations for systems of increasing length are performed by replicating the simulation cells along the temperature gradient direction, while maintaining the cross-sectional area of the simulation cell, it is possible that finite size effects are responsible for part of the difference.
To verify this, we also simulated models based on the 216-atom cell, with a smaller cross-sectional area equal to $(19$ \AA$)^2$.
For a cell with length 458.40 \AA~ along the heat-flux direction we found $\kappa(L)=0.211 \pm 0.006$ \wmk~, larger than our converged results but closer to the reported ab initio calculations.

\subsection{Homogeneous nonequilibrium molecular dynamics}

\begin{figure}[!htbp]
\includegraphics[width=\linewidth]{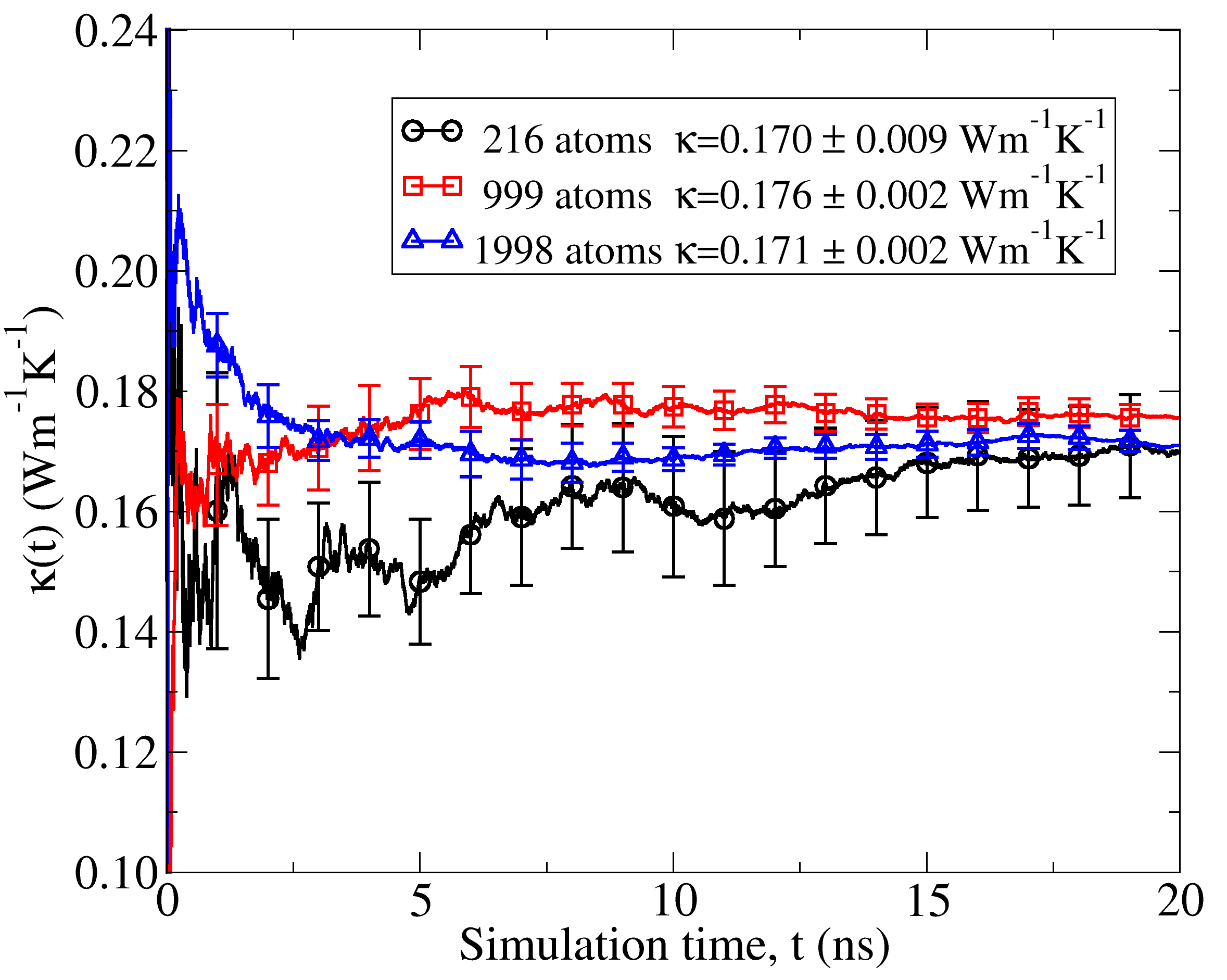}
\caption{Average running lattice thermal conductivity of aGST computed with HNEMD for 3 system sizes at $T=300$ K. The error bars represent the standard error within 10 realizations.}
\label{fig:hnemd}
\end{figure}

We further corroborate the results obtained with equilibrium and nonequilibrium MD, by performing the LTC calculation with the HNEMD method.
In the case of HNEMD, we determined that an external force parameter with magnitude $F_e=0.001$ \AA$^{-1}$ was sufficient to provide reliable results, after verifying that smaller and larger values yield consistent conductivities within the linear-response regime.
We considered three system sizes, corresponding to 216, 999 and 1998 atoms, and ran 10 independent simulations with a total time of 20 ns each.
Figure \ref{fig:hnemd} presents the running LTC averaged over 10 independent realizations, where the error bars represent the standard error of the mean considering all independent realizations.
After an initial transient period ($t < 5$ ns), the LTC converges to its stationary value, with a small size-dependence.

For the largest cell, with 1998 atoms, we obtain $\kappa_{\text{HNEMD}}=0.171 \pm 0.002$ \wmk, in agreement with the other MD results.
Running 10 $\times$ 20 ns HNEMD simulations on gaming GPUs (RTX3060/4070) is considerably faster than running 4 $\times$ 90 ns equilibrium MD simulations on 128-core CPUs (Intel Xeon Gold 6338), which is a practical advantage of the HNEMD method.
One disadvantage is that HNEMD is currently available only in GPUMD, which is not yet compatible with most machine-learned potentials besides NEP.

\subsection{Allen--Feldman}

The harmonic AF theory provides a disorder-limited reference for the LTC of aGST.
The difference between the LTC computed with MD and AF can then be related to anharmonic effects once classical specific heat effects are accounted for.
In crystals, a harmonic description cannot yield a finite conductivity, because anharmonic phonon-phonon scattering is the dominant resistive mechanism; anharmonicity therefore always reduces the conductivity with respect to its harmonic limit.
On the other hand, for amorphous materials the LTC is determined by a combination of structural disorder and anharmonicity.
In that case, when AF and MD are compared using the same classical specific heat, one can observe $\kappa_{\text{AF}} > \kappa_{\text{MD}}$, $\kappa_{\text{AF}} < \kappa_{\text{MD}}$, or even $\kappa_{\text{AF}} \approx \kappa_{\text{MD}}$.

\begin{figure}[!htbp]
\includegraphics[width=\linewidth]{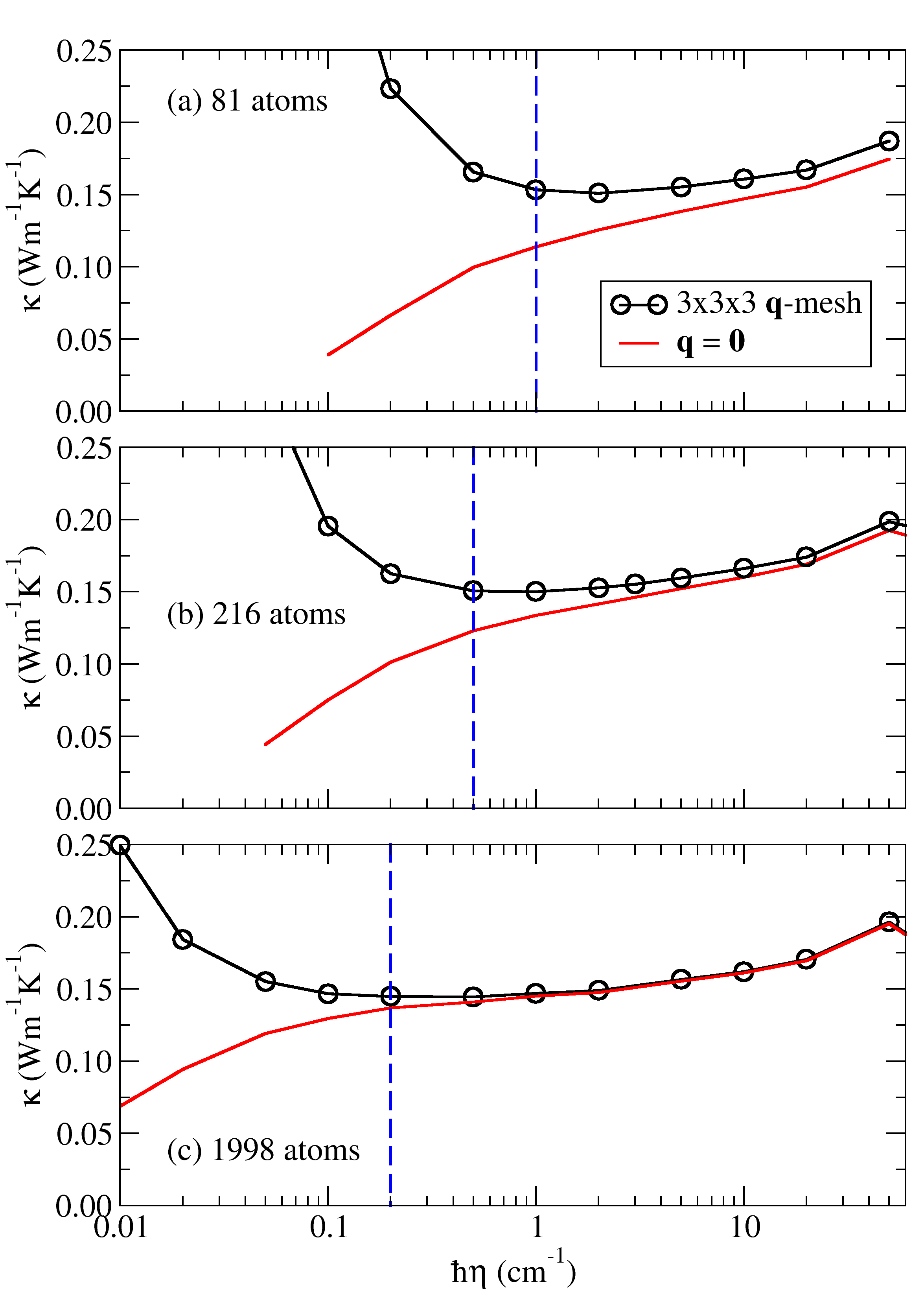}
\caption{Convergence of the Allen--Feldman conductivity as a function of the width of the delta function, calculated at $T=300$ K for $\bm{q}{=}\bm{0}$ and on a 3$\times$3$\times$3 $\bm{q}$-mesh. The dashed vertical lines indicate the beginning of the convergence plateau.}
\label{fig:k-vs-eta}
\end{figure}

The first step in the AF calculation is to determine an appropriate value for the finite width of the Dirac delta function, i.e. the broadening parameter $\eta$ \cite{Allen1989}.
In practice, if the system is large enough, there is a range of values of $\eta$ for which the LTC is essentially constant, a so-called convergence plateau \cite{Simoncelli2023, Harper2024, iwanowski_bond-network_2025}.
Figure \ref{fig:k-vs-eta} shows the dependence of $\kappa_{\text{AF}}$ on the broadening parameter $\eta$ for three system sizes computed at $\bm{q}{=}\bm{0}$ and on a 3$\times$3$\times$3 $\bm{q}$-mesh at $T=300$ K.
We have verified for all system sizes that a denser 5$\times$5$\times$5 $\bm{q}$-mesh does not significantly affect the numerical results.
The values computed on the $\bm{q}$-mesh demonstrate a clear convergence plateau, which grows with system size, and whose starting point is indicated by dashed vertical lines.
Nonetheless, the value of $\kappa_{\text{AF}}$ at the beginning of the convergence plateau shows small dependence on the system size: $\kappa_{\text{AF}}(N=81,\hbar \eta=1.0$ \cmm $)=0.154$ \wmk, $\kappa_{\text{AF}}(N=216,\hbar \eta=0.5$ \cmm $)=0.152$ \wmk, and $\kappa_{\text{AF}}(N=1998,\hbar \eta=0.2$ \cmm $)=0.144$ \wmk.

One feature in Fig.~\ref{fig:k-vs-eta} is the increase of $\kappa_\text{AF}$ for large $\eta$, which shows that finite-frequency velocity-operator couplings enhance the conductivity once they are sampled by a broader line shape.
This behavior is in contrast with what was observed in other amorphous materials such as silica and alumina, in which $\kappa_{\text{AF}}$ is constant or decreases for large $\eta$ \cite{Simoncelli2023, Harper2024}.
Furthermore, this trend is consistent with the relative difference between AF and MD, since $\kappa_{\text{MD}} > \kappa_{\text{AF}}$.
Considering the 1998-atom cell, for which both AF and HNEMD results are available, the ratio $\kappa_{\text{AF}}/\kappa_{\text{HNEMD}} \approx 0.84$ suggests that approximately $16$\% of the room-temperature LTC of aGST originates from effects absent in the harmonic AF limit.

\subsection{Wigner thermal conductivity} 

A full calculation of the LTC including anharmonic effects requires the computation of phonon linewidths, or analogously, their lifetimes at each temperature of interest.
We performed this computation considering three-phonon scattering processes within the RTA in phono3py \cite{Togo2023}, and included isotope scattering in the linewidths entering the rWTE.
The computation of phonon lifetimes is particularly demanding and it was performed on a 3$\times$3$\times$3 $\bm{q}$-mesh for the 81-atom cell, and only at $\bm{q}{=}\bm{0}$ for the 216-atom cell.
We then employed the recently developed interpolation scheme to obtain phonon lifetimes at arbitrary $\bm{q}$-points for the 216-atom cell \cite{Simoncelli2023}.

\begin{figure}[!htbp]
\includegraphics[width=\linewidth]{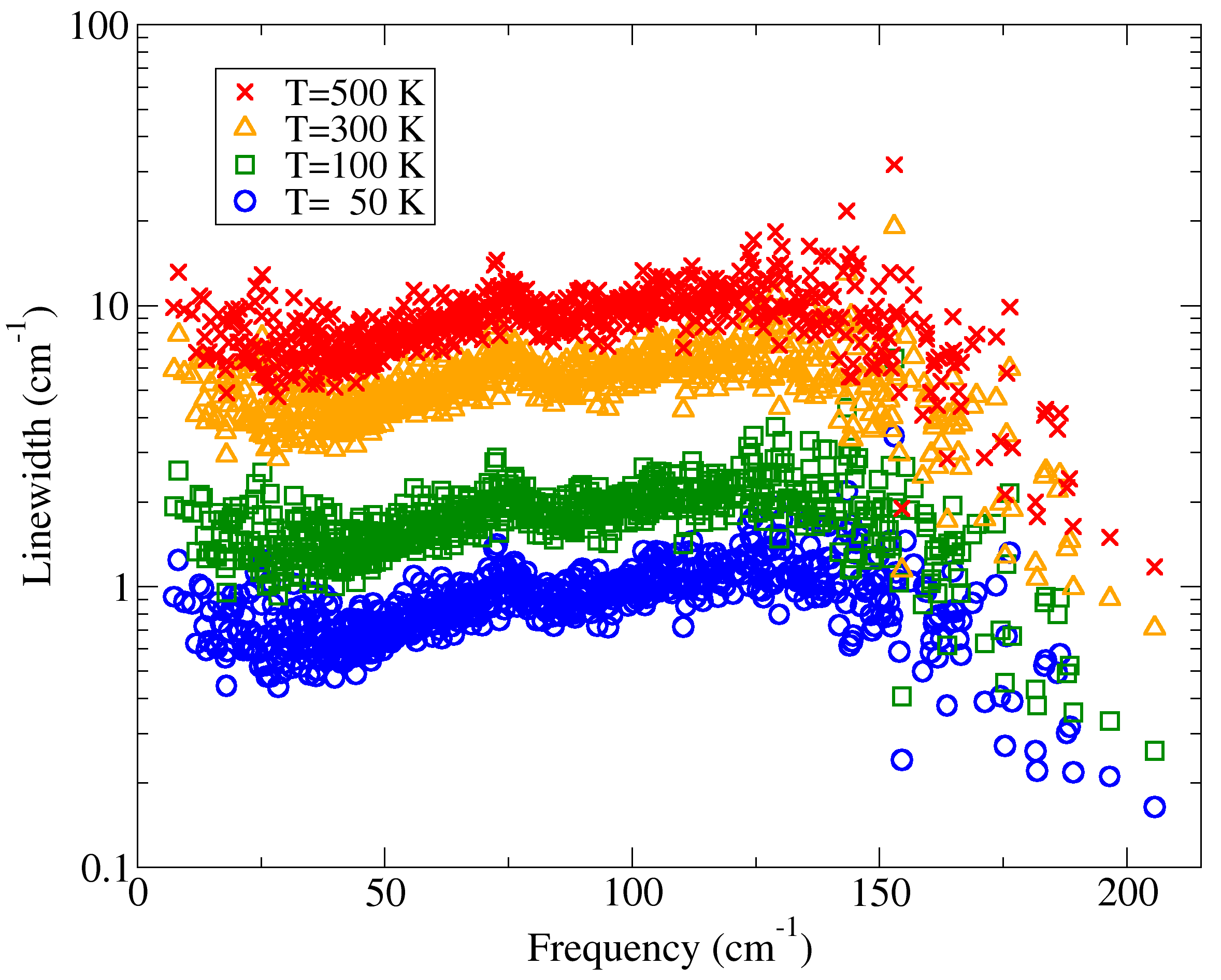}
\caption{Anharmonic linewidths considering only
three-phonon scatterings within the relaxation time approximation, computed at $\bm{q}{=}\bm{0}$ on a 216-atom cell at various temperatures.}
\label{fig:linewidth}
\end{figure}

Figure \ref{fig:linewidth} shows the linewidths computed with NEP at $T=$ 50, 100, 300, and 500 K.
Linewidths generally increase with temperature because of the increasing occupation of the vibrational modes, which enlarges the phase space for three-phonon scattering; the third-order force constants themselves are temperature independent in our calculations. 
In particular, the glass transition for aGST occurs around 450 K \cite{ElKheir2024}, but this transition is not sampled in the linewidth calculation, which evaluates vibrational properties around an amorphous reference structure.
 
\begin{figure}[!htbp]
\includegraphics[width=\linewidth]{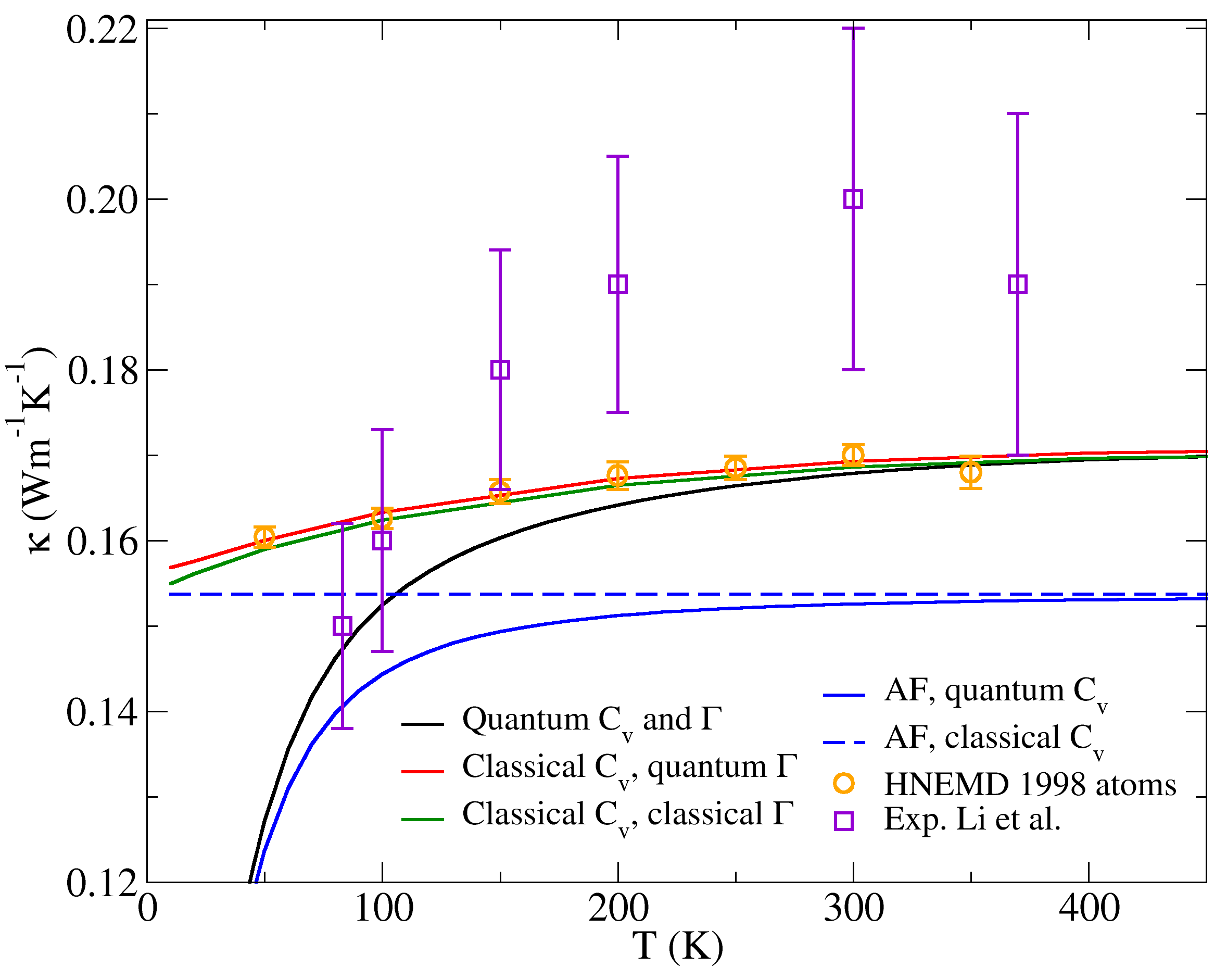}
\caption{Lattice thermal conductivity of aGST as a function of temperature with the Allen--Feldman model, the rWTE, and HNEMD. LD calculations were performed with the 216-atom cell, considering quantum and classical heat capacity and lifetimes. MD data are for the size-converged 1998-atom cell. Solid black, red, and green lines are rWTE results, and blue lines are AF results. Open squares are experimental data obtained by Li et al.~\cite{Li2023} with time-domain thermoreflectance on amorphous thin films.}
\label{fig:kappa-vs-T}
\end{figure}

The temperature dependence of the LTC of aGST computed with NEP is presented in Fig.~\ref{fig:kappa-vs-T}, considering the fully harmonic AF model, as well as the anharmonic rWTE.
In crystalline materials, the LTC usually decreases with temperature above $\Theta_{\rm Debye}$ because phonon-phonon scattering increases.
In amorphous materials, by contrast, the LTC can be nearly independent of $T$ when vibration-defect scattering is the dominant mechanism.
Although aGST can crystallize upon heating \cite{Li2023,ElKheir2024}, this phenomenon cannot be captured by the static LD calculations involved in the AF and rWTE methods.
However, in the HNEMD simulations, crystallization begins around 350 K, possibly promoted by the finite heat current.
Furthermore, the NEP potential was not specifically trained for crystallization, and therefore we do not use these simulations to quantify the crystallization process.

Several features are apparent in Fig.~\ref{fig:kappa-vs-T}.
First, the rWTE LTC is larger than the AF value.
This clearly indicates that in aGST anharmonic linewidths increase the LTC.
This conclusion is also consistent with the fact that the LTC computed from HNEMD is higher than the other two.
One main difference between our MD and LD calculations is that quantum effects are present only in the latter.
However, we can modify the calculation of the LTC and replace the quantum Bose--Einstein factors by their classical limit, either in the calculation of $C_v$, in the calculation of the linewidths, or in both.

Using the classical limit for $C_v$ only, or for both $C_v$ and the linewidths, yields larger LTC values that are compatible with the HNEMD results.
It is also possible to use the classical $C_v$ in the AF calculation, in which case the result becomes temperature-independent.
This raises the question of why aGST behaves differently from other amorphous materials, in the sense that its LTC increases when the material becomes more anharmonic upon increasing temperature.

The experimental thermal conductivity of aGST measured by Li et al. with time-domain thermoreflectance \cite{Li2023} is also shown in Fig.~\ref{fig:kappa-vs-T}. The measurements display a glass-like, monotonic increase of $\kappa$ with temperature below 300~K, which is best reproduced by the rWTE with quantum specific heat and linewidths; at 80--100~K the rWTE values fall within the experimental error bars, whereas the classical calculations, which lack the quantum reduction of the specific heat, lie systematically above the quantum ones and at the upper edge of the experimental error bars. Importantly, the measured conductivity keeps increasing between $\sim$150 and 300~K, where the quantum specific heat is nearly saturated and the harmonic AF conductivity is essentially flat. Within the experimental uncertainty, this increase is consistent with the anharmonic enhancement predicted by the rWTE and MD. At 300~K our calculations lie 10--15\% below the measured value of $0.20\pm0.02$~\wmk, at the lower edge of the experimental error bar. Possible sources of this residual difference include structural differences between the sputtered, as-deposited films and our melt-quenched models (density, degree of relaxation, and medium-range order) and the approximate description of interatomic bonding by the semilocal PBE functional, used without dispersion corrections in the DFT reference data~\cite{ElKheir2024}; the electronic contribution is negligible in the amorphous phase. The decrease of the highest-temperature experimental point is likely associated with structural relaxation preceding crystallization \cite{Li2023}, which is not captured by our calculations.

\section{Discussion}

In Table \ref{tab:results} we summarize our estimates for the LTC of aGST at $T=300$ K, obtained with MD- and LD-based methods with the NEP potential. 
There is a remarkable agreement among MD-based methods, regardless of their intrinsic differences, such as whether a thermostat is used and which one.
Although a precise computation of the uncertainty in $\kappa_\text{rWTE}$ is not available directly in our procedure, we find that the numerical value lies just below the lower limit computed from MD. Figure~\ref{fig:kappa-vs-T} shows that the difference with respect to HNEMD is largely explained by the use of quantum rather than classical specific heats in rWTE: with classical statistics, the rWTE value at 300~K becomes compatible with HNEMD.
The equilibrium-MD and RNEMD values remain about 5\% higher than the classical rWTE result. This residual difference may originate from anharmonic effects beyond third order, which are included in MD but not in the rWTE, from the RTA used for the linewidths, and from the different cell sizes and structural models used in the MD (1998 atoms) and LD (216 atoms) calculations.
As discussed above, all computed values lie slightly below the room-temperature experimental value of Li et al.~\cite{Li2023}, which is in turn consistent with the earlier 3$\omega$ result of $0.21\pm0.02$~\wmk~\cite{fallica}.

\newcolumntype{C}{>{\centering\arraybackslash}X}
\begin{table}[h]
\caption{Lattice thermal conductivity of aGST at $T=300$ K computed with MD- and LD-based methods, using NEP. MD results are for the 1998-atom cell, while LD (Allen--Feldman and rWTE) results are for the 216-atom cell.}
\label{tab:results}
\centering
\begin{tabularx}{\linewidth}{|C|C|C|}
\hline
  Method & Thermostat & $\kappa$ (\wmk) \\ 
\hline
\hline
    Green--Kubo & -- & $0.181 \pm 0.005$  \\
    Helfand--Einstein & -- & $0.178 \pm 0.005$  \\
    RNEMD & Nosé--Hoover & $0.180 \pm 0.005$  \\
    HNEMD & Bussi--Donadio--Parrinello & $0.171 \pm 0.002$  \\
\hline
    Allen--Feldman & -- & $0.152$  \\
\hline
    rWTE & -- & $0.168$  \\
\hline
\end{tabularx}
\end{table}

To identify the microscopic mechanism behind the anharmonicity-mediated increase of conductivity, we start from the Voigt distribution in Eq.~(\ref{eq:thermal_conductivity_combined}). In the high-temperature limit, the rWTE conductivity becomes a Lorentzian-weighted average of velocity-operator elements $|\!|\tens{v}(\bm{q})_{s,s'}|\!|^2\!$, with the largest weight carried by pairs satisfying $|\omega_{\bm{q}s}{-}\omega_{\bm{q}s'}|{<}[\Gamma_{\bm{q}s}{+}\Gamma_{\bm{q}s'}]$. Since the linewidths increase with temperature, progressively more pairs with finite frequency difference contribute to the average.
The temperature trend of the conductivity is therefore controlled by how $|\!|\tens{v}(\bm{q})_{s,s'}|\!|^2\!$ varies with the energy difference $\hbar|\omega_{\bm{q}s}{-}\omega_{\bm{q}s'}|$: if the velocity-operator elements increase (decrease) with frequency difference, the conductivity increases (decreases) with temperature. In the harmonic AF limit, by contrast, the conductivity is controlled only by quasi-degenerate couplings, $\hbar|\omega_{\bm{q}s}{-}\omega_{\bm{q}s'}|\to 0$, and depends on temperature only through the specific heat.
This criterion distinguishes different disorder regimes. In crystalline or weakly disordered solids, the velocity-operator elements empirically decrease with frequency difference\cite{simoncelli_temperature-invariant_2025}. In prototypical disordered solids such as vitreous silica, they are approximately independent of frequency difference above $\sim50$ cm$^{-1}$\cite{simoncelli_temperature-invariant_2025}. In aGST, in contrast, our results indicate that the effective velocity-operator couplings increase with frequency difference, explaining why anharmonic broadening enhances the conductivity.

\begin{figure}
\includegraphics[width=\linewidth]{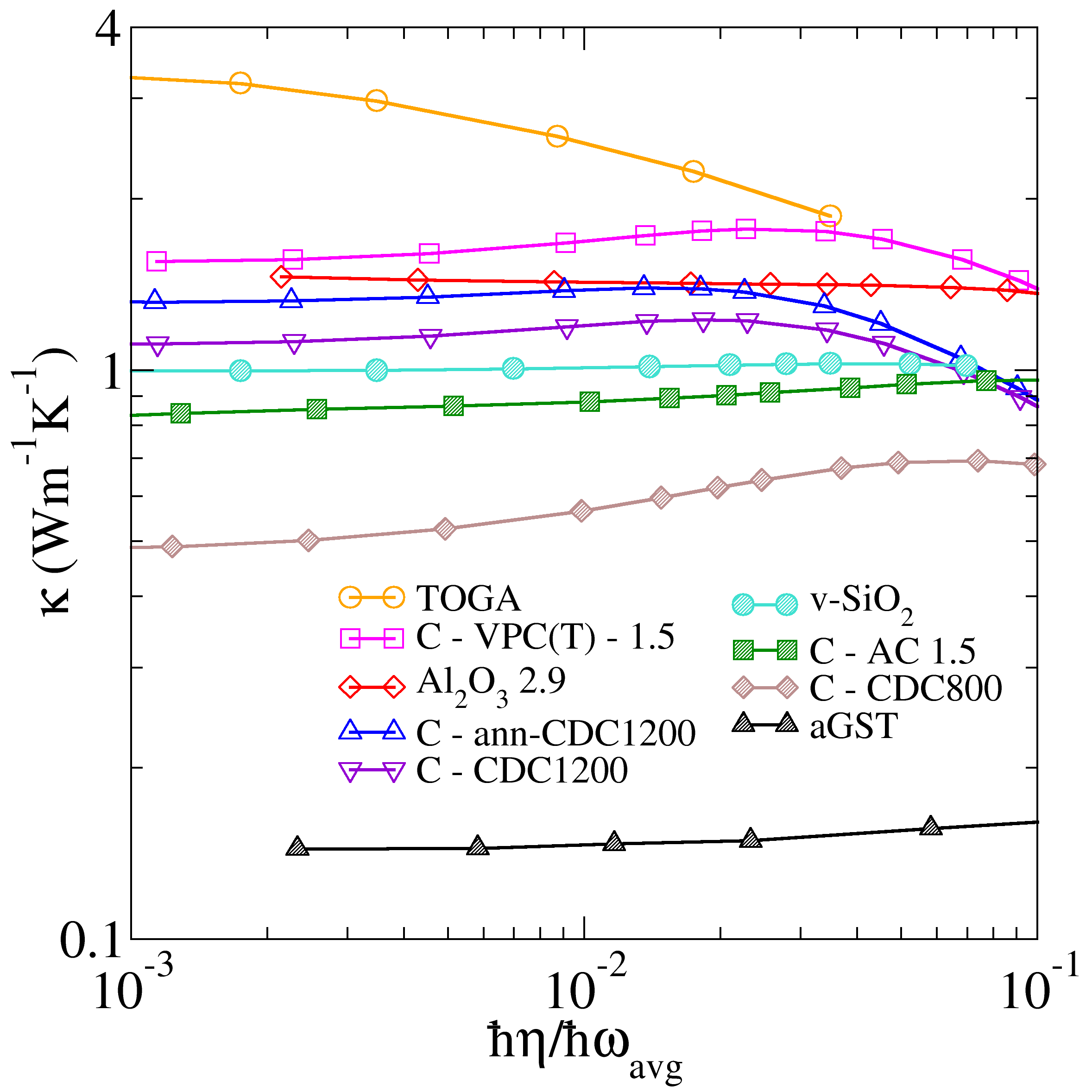}
\caption{Room-temperature lattice thermal conductivity of disordered silica, carbon, alumina and aGST as a function of the AF broadening $\eta$ divided by the average vibrational frequency $\omega_\text{avg}$. Direct calculations are shown as symbols, and lines show their smooth interpolations.
Legend labels: TOGA, topologically ordered geometrically amorphous silica \cite{simoncelli_temperature-invariant_2025};
v-SiO$_2$, vitreous silica \cite{Simoncelli2023}; Al$_2$O$_3$ 2.9, amorphous alumina at a density of 2.9~g/cm$^3$ \cite{Harper2024}; C -- AC 1.5, homogeneous amorphous carbon at 1.5~g/cm$^3$; C -- VPC(T) 1.5, nanoporous carbon at 1.5~g/cm$^3$; C -- CDC800 and C -- CDC1200, nanoporous carbide-derived carbons corresponding to synthesis temperatures of 800 and 1200~$^\circ$C, respectively; C -- ann-CDC1200, CDC1200 after further annealing; aGST, the 1998-atom model of this work. All carbon models are from Ref.~\cite{iwanowski_bond-network_2025}.
Gaussian smearing is used for the broadening of the delta function, and results are reported only in the domain of convergence where AF calculations using only the $\bm{q}=\bm{0}$ point and those using a $\bm{q}$-mesh differ negligibly.}
\label{fig:k-vs-etaomega}
\end{figure}
To test this interpretation and rationalize the difference between aGST and other amorphous materials, we present in Fig.~\ref{fig:k-vs-etaomega} the dependence of $\kappa$ on a normalized parameter $\hbar \eta / \hbar \omega_{\rm avg}$, where AF broadening $\eta$ is divided by the average vibrational frequency $\omega_\text{avg} = \left[ \int_0^{\omega_{\rm max}} d\omega \, \omega g(\omega) \right] /  \left[\int_0^{\omega_{\rm max}} d\omega \, g(\omega) \right]$ ($g(\omega)$ is the vibrational density of states). This normalization allows us to compare structures with different ranges of vibrational frequencies.
We analyze the changes in $\kappa$ for a wide range of materials including aGST, disordered carbon (amorphous AC 1.5, carbide-derived CDC800, CDC1200 and ann-CDC1200 \& nanoporous VPC(T) 1.5) \cite{iwanowski_bond-network_2025}, amorphous alumina \cite{Harper2024} and silica structures with different levels of disorder (TOGA, v-SiO$_2$) \cite{Simoncelli2023, simoncelli_temperature-invariant_2025}. 
We find three distinct behavior regimes.
First, conductivity approaches convergence with a negative derivative with respect to $\eta$, which is typical of pristine or weakly disordered crystals\cite{Fugallo2013,qianPhononengineeredExtremeThermal2021,dragasevicViscousHeatBackflow2026}.
In fact, for crystals, in the limit of $\eta \rightarrow 0$, conductivity diverges because $\eta$ acts as an artificial linewidth in the absence of anharmonicity, and conductivity scales as $1/\eta$.
We observe this decreasing trend in TOGA silica, which is less disordered than vitreous silica v-SiO$_2$\cite{simoncelli_temperature-invariant_2025}.
Second, conductivity reaches convergence with practically zero derivative, which is typical of standard glasses, such as amorphous silica\cite{Simoncelli2023} and amorphous alumina\cite{Harper2024}, and indicates a weak effect of anharmonicity at room temperature.
Third, conductivity approaches convergence with a positive derivative, as observed in aGST (see also Fig.~\ref{fig:k-vs-eta}) and in the heavily disordered nanoporous carbon CDC800\cite{iwanowski_bond-network_2025}.
This trend indicates that quasi-degenerate ($\omega_s \approx \omega_{s'}$) velocity-operator couplings are smaller than couplings between modes with nonzero frequency difference \cite{iwanowski_bond-network_2025}.

\section{Conclusions}

We have investigated the lattice thermal conductivity of amorphous Ge$_2$Sb$_2$Te$_5$ using equilibrium and nonequilibrium molecular dynamics, harmonic Allen--Feldman theory, and the anharmonic Wigner transport equation, all based on the same machine-learned interatomic potential.
The comparison separates the three physical ingredients that control heat transport in aGST: structural disorder, quantum vibrational statistics, and anharmonicity.
Allen--Feldman theory provides the harmonic, disorder-limited reference; the Wigner formulation adds quantum statistics and perturbative anharmonic linewidths; and molecular dynamics includes anharmonicity to all orders, but in the classical limit.

The molecular-dynamics methods yield mutually consistent room-temperature conductivities and are also in line with previous work based on \textit{ab initio} approach-to-equilibrium MD \cite{Mocanu2020}.
For the largest aGST models, the Green--Kubo, Helfand--Einstein, RNEMD, and HNEMD calculations yield values between $0.171$ and $0.181$ \wmk.
This agreement is significant because these approaches differ in their statistical estimators, boundary conditions, thermostats, and convergence requirements.
The harmonic Allen--Feldman conductivity, $\kappa_{\rm AF}=0.152$ \wmk, is lower than the molecular-dynamics values, while the regularized Wigner result, $\kappa_{\rm rWTE}=0.168$ \wmk, lies very close to the molecular-dynamics range.
When the Wigner calculation is evaluated with classical vibrational statistics, it agrees directly with HNEMD, and lies within about 5\% of the equilibrium and RNEMD values.
This shows that the difference between quantum Wigner transport and classical molecular dynamics originates mainly from their different statistical descriptions of the vibrational modes, rather than from an inconsistency between the transport formalisms.

Our results demonstrate that anharmonicity does not have a universal effect in disordered solids.
In crystalline or weakly disordered systems, anharmonicity usually limits heat conduction by increasing phonon scattering.
In standard amorphous solids such as vitreous silica and amorphous alumina, room-temperature transport is instead predominantly limited by structural disorder, and anharmonicity has a weak effect.
In strongly disordered and anharmonic aGST, the opposite behavior emerges: anharmonicity enhances thermal transport.
This conclusion is supported both by the fact that molecular dynamics and rWTE give conductivities larger than the harmonic Allen--Feldman value, and by the positive dependence of the conductivity on the normalized broadening parameter, which indicates that finite-frequency vibrational couplings enhance heat transport in aGST.
Thus aGST belongs to a regime where anharmonic linewidths activate additional couplings between vibrations, increasing the conductivity beyond its harmonic disorder-limited value.

We conclude that the Wigner transport equation provides a consistent bridge between harmonic Allen--Feldman theory and classical molecular dynamics for strongly disordered solids, provided that the statistical limit used in the comparison is specified.
For aGST, this bridge reveals that the low thermal conductivity is not simply the result of strong anharmonic scattering: it reflects the competition between structural disorder and anharmonic vibrational mixing.
This distinction is essential for phase-change materials and, more generally, for amorphous solids in which disorder and anharmonicity are both strong.

\begin{acknowledgments}
LFCP thanks Atsushi Togo for  assistance with phonopy and phono3py code modifications, and acknowledges financial support from CAPES (Grant 001), CNPq (Grants 406447/2022-5 and 371610/2023-0 INCT Materials Informatics, 408038/2022-5, and 310262/2025-9), FACEPE (Grant APQ-1117-1.05/22), and the visiting professors program at Sapienza Università di Roma. 
This work was also supported by the ICSC—Centro Nazionale di Ricerca in High Performance Computing, Big Data, and Quantum Computing funded by European Union—NextGenerationEU.
We also acknowledge the provision of high performance computational resources by CENAPAD-SP (proj968), the National Laboratory for Scientific Computing (LNCC/MCTI, Brazil - project nanophon on the Santos Dumont supercomputer), and Centro de Computação de Alto Desempenho (CCAD) at UFPE on its BoiVoador cluster.
\end{acknowledgments}

\section*{DATA AVAILABILITY}
The training DFT database is available in the Materials Cloud repository via https://doi.org/10.24435/materialscloud:a8-45.
Requests for further information or data should be sent to the authors.
LAMMPS, GPUMD, phonopy and phono3py are free and open source codes available at https://lammps.org, https://www.gpumd.org, https://phonopy.github.io/phonopy/ and https://phonopy.github.io/phono3py/, respectively.

\bibliographystyle{apsrev4-1}
\bibliography{library}

\end{document}